\documentclass[12pt]{article}
\usepackage{epsf}

\newcommand{\beq}{\begin{equation}}
\newcommand{\beql}[1]{\begin{equation}\label{eq:#1}}
\newcommand{\eeq}{\end{equation}}
\newcommand{\be}{\begin{equation}}
\newcommand{\ee}{\end{equation}}
\newcommand{\beqn}{\begin{eqnarray}}
\newcommand{\eeqn}{\end{eqnarray}}
\newcommand{\bea}{\begin{eqnarray}}
\newcommand{\eea}{\end{eqnarray}}

\newcommand{\eq}[1]{(\ref{eq:#1})}

\DeclareFixedFont{\xiiss}{OT1}{cmss}{m}{n}{12}
\DeclareFixedFont{\ixss}{OT1}{cmss}{m}{n}{9}
\DeclareFixedFont{\cmrnine}{OT1}{cmr}{m}{n}{9}

\newcommand{\CC}{\hbox{\xiiss C\kern-.4emI}}
\newcommand{\RR}{\hbox{\xiiss R\kern-.45emI}}
\newcommand{\PP}{\hbox{\xiiss P\kern-.45emI}}
\newcommand{\ZZ}{\hbox{\xiiss Z\kern-.4emZ}}
\newcommand{\RRs}{\hbox{\ixss R\kern-.5emI}}
\newcommand{\CCs}{\hbox{\ixss C\kern-.4emI}}
\newcommand{\ZZs}{\hbox{\ixss Z\kern-.4emZ}}

\newcommand{\tr}{{\rm tr}\ }
\newcommand{\tpi}{2\pi i}

\newcommand{\myfig}[3]{\begin{figure}[ht]
\begin{center}
\leavevmode
\epsfxsize=#2cm
\epsfbox{#1.EPSF}
\end{center}
\caption{#3}
\label{fig:#1}
\end{figure}}

\begin{document}
        \begin{titlepage}
        \title{
                \begin{flushright}
                \begin{small}
                ILL-(TH)-98-04\\
                hep-th/9807082\\
                \end{small}
                \end{flushright}
%               \vspace{1.cm}
%          Consistency~Conditions of  Brane~Box Models
		Brane~Boxes,~Anomalies,~Bending~and~Tadpoles
        }

        \author{
        Robert G. Leigh\thanks{e-mail: \tt rgleigh@uiuc.edu}
        \\
        \ and\\
        Moshe Rozali\thanks{e-mail: \tt rozali@pobox.hep.uiuc.edu}\\
        \\
                {\small\it Department of Physics}\\
                {\small\it University of Illinois at Urbana-Champaign}\\
                {\small\it Urbana, IL 61801}\\
        }

        \maketitle

        \begin{abstract}

Certain classes of chiral four-dimensional gauge theories may be
obtained as the worldvolume theories of D5-branes suspended between
networks of NS5-branes, the so-called brane box models. In this paper,
we derive the stringy consistency conditions placed on these models, and
show that they are equivalent to anomaly cancellation of the gauge
theories. We derive these conditions in the orbifold theories which are
T-dual to the elliptic brane box models. Specifically, we show that the
expression for tadpoles for unphysical twisted Ramond-Ramond 4-form
fields in the orbifold theory are proportional to the gauge anomalies of
the brane box theory. Thus string consistency is equivalent to
worldvolume gauge anomaly cancellation. Furthermore, we find additional
cylinder amplitudes which give the $\beta$-functions of the
gauge theory. We show how these correspond to bending of the NS-branes
in the brane box theory.
        \end{abstract}

        \end{titlepage}

%%%%%%%%%%%%%%%%%%%%%%%%%%%%%%%%%%%%%%%%%%%%%%%%%%%%%%%%%%%%%%%%%%%%%

\section{Introduction}

Following the work of \cite{hw}, the study of brane configurations in
string theory and M-theory has provided a useful description
of supersymmetric field theories.\footnote{For a recent review and
references see \cite{rev}.} Given the success of this approach, it is
desirable to utilize it to confront the interesting questions of four
dimensional chiral gauge theories with $ N =1$ SUSY, primarily the
question of spontaneous supersymmetry breaking.  Certain chiral theories have
been constructed in \cite{chiral}, but a general construction has proven
to be difficult.

Such a  construction of a fairly general class of  chiral gauge theories
has been introduced in Ref. \cite{hanzaf}. The simplest such models
(with no orientifold planes) have a  product of unitary gauge groups and
bifundamental matter. They are constructed as the worldvolume theory of
D5 branes that extend in 4 non-compact directions, and are finite in two
directions (denoted 4,5). In the directions (4,5) the D5 branes end on
infinite NS fivebranes. Such a model is depicted in Figure 1; horizontal
and vertical lines denote NS5-branes, while shaded areas represent
D5-branes, of which there are $N_{rs}$ in the $(r,s)^{th}$ box. Arrows
between boxes represent chiral multiplets, whose $SU(N)$ representation
is by convention fundamental for outgoing arrows. A precise  description
of the matter content and the matter  couplings can be found in
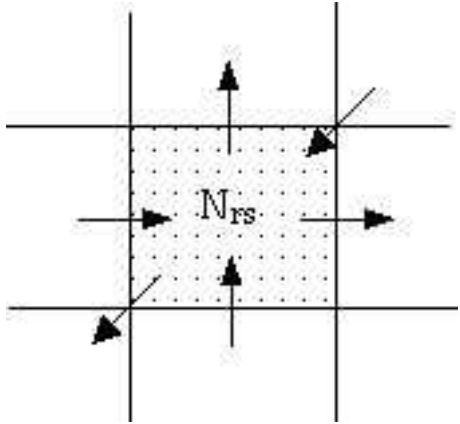
\cite{hanzaf}. \myfig{simplebox}{6}{One box of a brane box model,
showing chiral multiplets.}

As given, the brane box represents the configuration at zero string
coupling. At a finite string coupling, it is an interesting question as
to which brane boxes (given by the ``rank matrix'' $\{ N_{r,s}\}$) are
consistent. Certainly, we expect that a necessary condition, at the level
of the field theory, is  gauge anomaly freedom. It is not immediately clear
how to derive this condition from string theory. The most imposing problem
is the complicated background implied by the presence of the NS5-branes, as 
well as the fact that $D5$-branes end on them.

Attempts at answering this question have been made. At finite coupling,
the NS5-branes generically will bend. Gimon and Gremm\cite{gimon} have
given a condition which they claim was necessary to avoid tearing of the
brane configuration. This suggestion has several problems; first, it
rules out several clearly consistent models, and second, it has been
recently shown\cite{lisa} that tearing does not occur, as there are
sufficient dimensions available into which the branes may distort. Brane
bending is in fact connected directly to renormalization group flow
(scale anomalies) rather than gauge anomalies.

The question remains then what exactly are the conditions for
consistency, and how they are understood in terms of string theory. In
this paper, we answer this question in the  case of ``elliptic'' models,
where identifications are made across the brane box (i.e. the system is
compactified in the 4,5 directions).  Since any brane box configuration
can be embedded in a periodic configuration (allowing for empty rows and
columns), it would seem that this causes no loss of generality. However,
since the directions 4,5 are taken to be finite, no gauge group factor
can be truly reduced to a flavor symmetry. Therefore one cannot realize
gauge theories with anomalous flavor symmetries in the elliptic
construction.\footnote{At least without introducing additional branes.}

In the elliptic cases, there is a T-dual configuration
\cite{hanura,karch} of D3-branes at an  orbifold singularity. The action
of the orbifold group on Chan-Paton factors is given in terms of the
rank matrix $N_{r,s}$. In this orbifold theory, it is a straightforward
application of previous work \cite{tadpole} to compute the consistency
conditions. We do this by computing certain  cylinder amplitudes, and
find generically that there are tadpoles for unphysical twisted
Ramond-Ramond 4-form potentials. Consistency of string theory in the
orbifold background requires the vanishing of all such tadpoles. We show
that these conditions are exactly the anomaly freedom conditions in the
gauge theory.

There are certain twist sectors in the cylinder amplitudes which do not
contribute to the unphysical tadpole. Instead, they lead to 
logarithmic behaviour for certain physical modes. The coefficient of
this is proportional to the $\beta$-function coefficients of the gauge theory. 
By identifying these modes and their T-dual partners, it is possible to
relate this to brane bending, at least in simple cases where the geometry is
most clear.

The paper is organized as follows. We begin by discussing anomalies in
the brane configuration from the field theory point of view. In
particular we derive the set of conditions that is necessary and
sufficient to ensure anomaly freedom in the gauge theory. We find the
general solution of these conditions for ``non-degenerate'' models
(defined to be boxes such that $N_{r,s} \neq 0 $ for all $r,s$). 
In
section 3 we discuss the consistency conditions for string theory in the
orbifold background, and show that these conditions are identical to
those derived from the field theory perspective. In Section 4, we discuss
the $\beta$-functions.

%%%%%%%%%%%%%%%%%%%%%%%%%%%%%%%%%%%%%%%%%%%%%%%%%%%%%%%%%%%%%%%%%%%%%

\section{Field Theory Analysis}

\subsection{ Non-Degenerate Cases}

We begin the discussion by identifying scale and gauge anomalies in
the worldvolume theory of a given brane box model. There will be one
anomaly for each $SU(N_{rs})$ factor. The gauge anomaly is  proportional
to
\beql{anomalies}
{\cal A}_{(r,s)}={1\over2}\big(
N_{r-1,s}+N_{r,s+1}+N_{r+1,s-1}-N_{r+1,s}-N_{r,s-1}-N_{r-1,s+1}\big).
\eeq
In a  consistent gauge theory  each of these expressions has to  vanish,
as long as $N_{r,s}\geq 1$. Obviously, for $N_{r,s}=0$, there is no
corresponding gauge group, and from the field theory perspective there
is no resulting constraint. The situation for $N_{r,s}=1 $ is more
surprising: our results indicate that even though the $U(1)$ theory
decouples in the infra-red, the anomaly constraints of this gauge theory
must be imposed for consistency of the string theory.

 We start by analyzing configurations for which $N_{r,s} \neq 0$
for all $r,s$; these configurations we will call ``non-degenerate.''  Later
on we extend the analysis to the  degenerate cases. In the non-degenerate cases
one imposes ${\cal A}_{r,s} = 0$ for all $r,s$.

The key to comparing these expressions to the tadpoles in an orbifold
realization of the gauge theory is to express them in terms of
Chan-Paton matrices which represent the orbifold group action. We do
this explicitly for the case of a $\ZZ_k\times \ZZ_{k'}$ orbifold. This
corresponds to a brane box with trivial identifications. The
generalization to the other cases is straightforward.

The correspondence is as follows. An irreducible  representation of $\ZZ_k\times
\ZZ_{k'}$
assigns to the two generators $ \alpha, \beta$ the phases 
$ \left( \omega_k^r,\omega_{k'}^s\right)\equiv
\left( e^{\tpi r/k}, e^{\tpi s/k'}\right)$. The 
Chan-Paton matrix representing the element $ {\alpha}^n {\beta}^{n'} $
is then 
a diagonal matrix $\gamma_{(n,n')}$, with $N_{r,s}$ entries of
$\omega_k^{nr}\omega_{k'}^{n's}$. Thus the Chan-Paton matrix
satisfies
\beq
\tr \gamma_{(n,n')} = \sum_{r,s} N_{r,s}\ \omega_k^{rn}\ \omega_{k'}^{sn'}
\eeq
Note then that T-duality can be thought of as a discrete Fourier transform
between the rank matrix $N_{r,s}$ and the Chan-Paton matrix $\gamma_{(n,n')}$, 
where twist plays the role of momentum. 
 In particular  $\tr\gamma_{(0,0)}=\tr 1=\sum_{r,s} N_{r,s}=n_3$,
where $n_3$ is the number of D3-branes in the orbifold theory.

Now, we can isolate  $N_{r,s}$  by an inverse Fourier transform:
\beql{fourn}
N_{r,s}={1\over kk'}\sum_{n,n'} \omega_k^{-nr} \omega_{k'}^{-n's}\
\tr\gamma_{(n,n')}.
\eeq
 Using this expression we can rewrite eq. \eq{anomalies}
\beql{foura}
{\cal A}_{(r,s)}=
{i\over kk'} \sum_{n,n'} \omega_k^{-nr} \omega_{k'}^{-n's}\
a_{n,n'}\ \tr\gamma_{(n,n')}
\eeq
where
\beqn
a_{n,n'}&=&\sin(2\pi z)-\sin(2\pi z')-\sin(2\pi (z-z'))\\
&\equiv& -4\sin\pi z\ \sin\pi z'\ \sin\pi(z-z')\nonumber
\eeqn
and  we have introduced the notation $z=n/k$ and $z'=n'/k'$.

Since the above expressions involve a sum over a complete set of Fourier
modes, the vanishing of the anomaly requires a vanishing of each
coefficient separately. Therefore  in an anomaly free theory we must
have either $a_{n,n'}=0$ or $\tr\gamma_{n,n'}=0$, for every $n,n'$.
These are  at most $kk'$ conditions on the rank matrix $N_{r,s}$, which
has $kk'$ entries. For every instance where $a_{n,n'}=0$ we expect then
to find a free integer parameter in the solution for $N_{r,s}$. 

Similarly, the one-loop coefficient of the $\beta$-function (scale anomaly) 
is 
\beql{betafn}
\beta_{(r,s)}=3N_{r,s}-{1\over2}\left(N_{r-1,s}+N_{r,s+1}+N_{r+1,s-1}+
N_{r+1,s}+N_{r,s-1}+N_{r-1,s+1}\right)
\eeq
These may be re-written as
\beql{betafnrs}
\beta_{(r,s)}=
{1\over kk'} \sum_{n,n'} \omega_k^{-nr} \omega_{k'}^{-n's}\
b_{n,n'}\ \tr\gamma_{(n,n')}
\eeq
where
\beqn\label{eq:betacoeff}
b_{n,n'}&=&3-\cos(2\pi z)-\cos(2\pi z')-\cos(2\pi (z-z'))\\
&=&-2\left( \sin^2\pi z +\sin^2\pi z' +\sin^2\pi(z-z')\right)\nonumber
\eeqn

We first comment on the general conditions for scale invariance of the
gauge theory. We note that $b_{0,0} =0$, and therefore there is no
restriction on $\tr\gamma_{0,0} = n_3$. However, all other traces must
vanish since $b_{n,n'}$ is non-zero. This singles out the regular
representation of  $\ZZ_k\times \ZZ_{k'}$, or any number of copies of
it, as the most general solution. The corresponding brane box model has
$N_{r,s} = N$ for every $r,s$. The free parameter here is the rank $N$
of each gauge group factor. Such models were considered in
\cite{hanmatt} and were shown to be finite theories. The vanishing of
the beta function was related to the absence of bending in the brane box
construction.

More generally, we now discuss the conditions for anomaly freedom of the
non-degenerate models. Note that $a_{n,n'} =0$  for $z=0$ , $z'=0$ or
$z=z'$ (the latter is only possible when $k,k'$ are not relatively
prime). Therefore there is no restriction on $\tr\gamma_{(n,n')}$ in
those cases. For all other values of $n,n'$ the corresponding trace has
to vanish. 

The most general solution to this condition  has a simple geometrical
interpretation in the brane box picture. We start with the models with
$\tr\gamma_{(n,0)}\neq 0$ and all other traces vanishing. The
corresponding box configurations are translationally invariant in the
vertical direction, and have $N =2$ SUSY. Similarly the two other
cases have horizontal and diagonal symmetry, and have therefore the
matter content and interactions of $N =2$ SUSY theory. The
horizontal and vertical models have $k',k$ free integers in $N_{r,s}$,
whereas the diagonal ones have $gcd(k,k') -1$ such free parameters.

Since the condition for anomaly freedom is linear in the rank matrix
$N_{r,s}$, one can superimpose configurations which are separately
non-degenerate. This process was called ''sewing'' in \cite{hanura}. We 
therefore find that the general class of ``sewn $N =2 $'' models
of \cite{hanura} is the general solution of the anomaly constraints for
non-degenerate models. This class of models has $k+k'+gcd(k,k')-2$ free integer
parameters in the matrix $N_{r,s}$. It includes (but is not restricted to)
the general solution to the conditions suggested by Gimon and Gremm\cite{gimon}.

We note that such ``sewing'' will in general give an anomalous theory if
one of the ingredients is a degenerate configuration. Such a configuration
is not required to satisfy ${\cal A}_{r,s} =0$ for all boxes, and therefore
can violate the necessary anomaly constraints once superimposed on a
non-degenerate configuration.

\subsection{General Case}

It is obvious that the above conditions do not yield the most general
anomaly free model. An obvious exception is the pure SYM theory that is
realized by a box configuration with a single non-vanishing entry in the
rank matrix $N_{r,s}$. To allow for such degenerate cases as well we
define the quantity:
\beq
C_{r,s} \equiv N_{r,s} {\cal A}_{r,s} \quad \mbox{(no summation) }
\eeq
 
Clearly, the vanishing of $C_{r,s}$ is exactly equivalent to anomaly
freedom of the corresponding gauge theory. This vanishing condition
reduces to ${\cal A}_{r,s} =0$ for the non-degenerate cases discussed above. We
note that this condition is no longer linear in the rank matrix
$N_{r,s}$, thus ``sewing'' models is not allowed in the general case, as
expected.

In the next section we derive this general condition from demanding
consistency of the string background, which requires the cancellation of
certain tadpoles in the orbifold background. To make connection to the
tadpole calculation it is convenient to define the quantity
\beq
C_{n,n'}\equiv \sum_{r,s} \omega_k^{nr} \omega_{k'}^{n's}\ N_{r,s}\ 
{\cal A}_{(r,s)}
\eeq
 which is simply the Fourier transform of $C_{r,s}$.
Using eqs. \eq{fourn}, \eq{foura}, we can then deduce
\beq
C_{n,n'}={i\over kk'}\sum_{m,m'}\ a_{m,m'} \
\tr\gamma_{m,m'}\ \tr\gamma^{-1}_{m-n,m'-n'}
\eeq

The anomaly freedom is thus equivalent to the vanishing of these
Fourier modes for every $n,n'$. This imposes at most $kk'$ conditions 
on the rank matrix $N_{r,s}$. 

We now turn to derive these conditions as consistency conditions of the
orbifold theory. The quantities $C_{n,n'}$ are shown in the next
section to correspond to certain cylinder amplitudes that have to
vanish to ensure background consistency.

\section{Tadpoles in the Orbifold picture}

Next we consider the T-dual orbifold in detail. This consists of $n_3$
D3-branes at a $\ZZ_k\times\ZZ_{k'}$ orbifold singularity. We choose
coordinates such that the orbifold action is generated by
\beqn
\alpha &:& (z_1,z_2,z_3) \to (\omega_k z_1,\omega_k^{-1} z_2, z_3)\\
\beta &:& (z_1,z_2,z_3) \to (z_1,\omega_{k'} z_2, \omega_{k'}^{-1} z_3)
\eeqn
Tadpoles may be deduced by factorization on the closed string channel
of the cylinder amplitude
\beq
\tr_{NS-R}\ {1\over2}\sum_{n=0}^{k-1}\sum_{n'=0}^{k'-1}\
{{\alpha}^n\beta^{n'}
\over kk'}\ {1+(-)^F\over2}\ e^{-2\pi tL_o}
\eeq
The amplitude may be deduced by study of Refs.
\cite{tadpole}.
Note that our computation is particularly straightforward: there is only
one worldsheet topology to consider, and there are only D3-branes.
Using notation given in \cite{tadpole},
we can write the full amplitude in terms of $\theta$-functions\footnote{We
simplify notation as follows: $\theta_k(z)\equiv \theta_k(z|t)$, defined
in the usual way in terms of $q=e^{-\pi t}$.} as follows:\footnote{We have
defined $v_4=V_4/(4\pi^2\alpha')^2$, where $V_4$ is the regulated
volume of spacetime.}
\beqn
A_{0,0'}&=&-{v_4\over 2kk'}\int {dt\over t} (2t)^{-2}\sum_{n,n'}
W_{n,n'}\ a_{n,n'}\ \tr\gamma_{(n,n')}
\tr\gamma_{(n,n')}^{-1}\\
&\times &\left[ {\theta_3(0)\theta_3(z)\theta_3(z-z')\theta_3(z')
-(3\leftrightarrow 2)
-(3\leftrightarrow 4)\over
f_1^3(t) \theta_1(z)\theta_1(z-z')\theta_1(z')}\right]
\nonumber
\eeqn
 The factor of $a_{n,n'}$
has appeared in the amplitude through the definition of the $\theta_1$'s.
The factor of $W_{n,n'}$ denotes the partition function of winding modes
which may be present in some twist sectors (when $z=0$, or $z'=0$ or
$z=z'$).

The quantity on the second line of course vanishes (abstrusely). The $t\to 0$
limit corresponds to a long cylinder, and thus the amplitude will factorize
into a product of tadpoles. In particular, there are tadpoles for Ramond-Ramond
forms which are unphysical; in a consistent theory, these must all vanish 
separately.
The expression above implies that in sectors $m,m'\neq 0$, for which $W_{n,n'}=1$ (no
winding modes), there is a tadpole for twisted RR 4-form potentials. Indeed,
in the factorization limit ($\ell= 1/t \to 0$), those sectors give\footnote{Purely
numerical overall factors will be dropped.}
\beq
A^{(4)}_{0,0'}\sim (1-1)\int d\ell\ V_4\ {i\over kk'}\sum_{m,m'\neq 0}\ a_{m,m'} \
\tr\gamma_{m,m'}\ \tr\gamma^{-1}_{m,m'} 
\eeq
This tadpole must be cancelled since the field is supported in the orbifold
singularity; there are no transverse directions to propagate in.

There are $kk'$ tadpoles that must be cancelled. One could 
isolate the tadpoles themselves, but it is more convenient to do the following
formal trick to isolate the $kk'$ quantities in a convenient fashion. Define
additional cylinder amplitudes, with the insertion of a twist operator,
as indicated in Figure 2. 
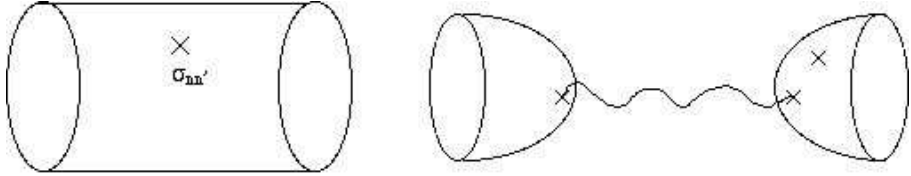
\myfig{factor}{12}{Factorization of cylinder, with a twist insertion.}
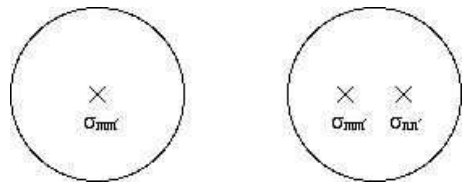
\myfig{discs}{6}{The corresponding disc amplitudes, appearing in the factorization.}
This corresponds to an injection of twist
"momentum" into the vacuum-vacuum amplitude, and in the 
factorization limit, we  obtain
\beq
A^{(4)}_{n,n'}=(1-1)\int d\ell\ V_4\ {i\over kk'}\sum_{m,m'\neq 0}\ a_{m,m'} \
\tr\gamma_{m,m'}\ \tr\gamma^{-1}_{m-n,m'-n'} 
\eeq
We obtain $kk'$ amplitudes which are bilinear in the $kk'$ tadpoles and therefore
must vanish. Note that these amplitudes are precisely the quantities $C_{n,n'}$,
defined in the last section in terms of $N_{r,s}$ and the field theory
anomalies ${\cal A}_{r,s}$. 

Thus we conclude that the cancellation of unphysical tadpoles
is precisely equivalent to anomaly cancellation. Note as well that the string theory
is smart enough to require cancellation of only those anomalies ($N_{r,s}\geq 1$)
which must be cancelled, again, because the amplitude is proportional to $N_{r,s}
{\cal A}_{r,s}$.

\section{The $\beta$-Functions}

We have seen that the twist sectors for which $m,m'\neq 0$ give rise to
unphysical tadpoles. There are also the twist sectors with $z=0$, $z'=0$
or $z=z'$, for which $W_{n,n'}\neq 1$. These will produce tadpoles for
additional, partially twisted, RR potentials. These sectors are special,
in that they effectively reduce to an A-type orbifold singularity.
Consequently, in the four-dimensional theory there is effectively $N=2$
supersymmetry, when restricted to one of these sectors.

The cylinder amplitude for these sectors is of the form:\footnote{We
continue to write a contribution from the $z=z'$ sector; as noted, it
exists only when $k$ and $k'$ are not relatively prime.}
\beqn
A^{(6)}_{0,0'}&=&(1-1)\int {d\ell\over\ell}\ {i\over kk'}\ v_4 \\
&&\times \left( \sin^2\pi z\ W_{z,0}\ \tr\gamma_{z,0}\ \tr\gamma^{-1}_{z,0}
+ \sin^2\pi z'\ W_{0,z'}\ \tr\gamma_{0,z'}\ \tr\gamma^{-1}_{0,z'}\right.\nonumber\\
&&\left.  \ \ \ + \sin^2\pi z\ W_{z,z}\ \tr\gamma_{z,z}\ \tr\gamma^{-1}_{z,z}\right)\nonumber
\eeqn
We have not yet evaluated the winding factors. They are, for example, of
the form:
\beq
W_{z,0}=\prod_{i=8,9}\sum_w e^{-2\pi t w^2 R_i^2}
\eeq
Strictly speaking, we are working at an
isolated orbifold point, and it is necessary to take the $v\to\infty$ limit.
This should be done before taking the $\ell\to\infty$ limit (and hence
Poisson resummation is unnecessary). Thus, at infinite volume, we get
\beqn\label{eq:phystad}
A^{(6)}_{0,0'}&=&{iv_4\over kk'}\ (1-1)\int {d\ell\over\ell}\\
&&\times \left( \sin^2\pi z\ \tr\gamma_{z,0}\ \tr\gamma^{-1}_{z,0}
+ \sin^2\pi z'\ \tr\gamma_{0,z'}\ \tr\gamma^{-1}_{0,z'}
+ \sin^2\pi z\ \tr\gamma_{z,z}\ \tr\gamma^{-1}_{z,z}\right)\nonumber
\eeqn
We now compare this expression with the $\beta$-function of the gauge theory,
\eq{betafnrs},\eq{betacoeff} , or more precisely to $N_{r,s}\beta_{r,s}$.
Given cancellation of unphysical tadpoles, the expression for $N_{r,s}\beta_{r,s}$
simplifies dramatically. As a result, we
find
\beql{phystada}
A^{(6)}_{0,0'}=v_4\ (1-1)\int {d\ell\over\ell}\sum_{r,s} N_{r,s}\beta_{r,s}
\eeq
As in the previous case, one could put in a twist in the cylinder amplitude
to reproduce all Fourier modes of $N_{r,s}\beta_{r,s}$. 

The logarithmic divergence $d\ell\over\ell$ in eq. \eq{phystada} has a clear
physical interpretation. Indeed, the propagator of a two-dimensional
field of mass $\epsilon$, at zero spacetime separation, may be written as
\beq
\int {dt\over t}e^{-t\epsilon^2}.
\eeq
Therefore we interpret $\epsilon$ as an infrared cutoff in
spacetime.\footnote{There is a similar discussion in Ref. \cite{mrdli}
with $N=2$ supersymmetry. In that case, one could regulate by moving on
the Coulomb branch (separating the 3-branes). In this $N=1$ case, the
3-branes cannot be moved off of the orbifold. The precise regularization
is immaterial to our discussion.} The divergence is caused by the propagation
of the partially twisted forms in two transverse directions.

We wish to absorb this divergence in the redefinition
of the couplings of the gauge theory on the D3-branes, in the general
spirit of renormalization theory, similar to the Fischler-Susskind
mechanism\cite{fissus}.  The general Wess-Zumino action
of the gauge theory is: \cite{wz}
\beq
S= \sum_{r,s}\int C^{(r,s)}\wedge Tr(e^{F_{r,s}})
 \eeq 
where $F_{r,s}$  is the field strength of $SU(N_{r,s})$, and $C^{(r,s)}$ is
a formal sum of couplings related to background values of RR forms in
spacetime.\footnote{ Specifically, $C^{(r,s)}$ is the Fourier transform
of $C_{n,n'}$, the background value of the RR form in the $(n,n')$
twisted sector.} The trace is taken in the fundamental representation
\cite{wz}.

Working in a particular  $(r,s)$ sector, the first term in this coupling
is:
\beql{wz}
\int C_{(4)}^{(r,s)} Tr (1) = \int C_{(4)}^{(r,s)} N_{r,s} .
\eeq
We see therefore that the regulated amplitude leads to a renormalization
of the coupling $C_{(4)}^{(r,s)}$ as given by
\beq
\delta C_{(4)}^{(r,s)} \sim  \beta_{r,s} \ln\epsilon  .
\eeq

Another coupling in \eq{wz} is the term $C_{(0)}\wedge F\wedge F$ , with
the same coefficient (since the couplings are related by T-duality).
This identifies $C_{(0)}^{(r,s)}$ as the gauge coupling of the $
{(r,s)}^{th}$ gauge group. We conclude therefore that the cylinder
amplitudes presented above indeed reproduce the correct running of the
gauge couplings.

One can demonstrate this more explicitly by considering the two-point
function of $F$ on the cylinder. It is convenient to perform this
calculation instead as the vacuum-vacuum cylinder diagram with a
background $F$ turned on. The $F^2$ term in the effective action is then
obtained by expansion for small $F$ \cite{bachas}. It is sufficient to
consider the cylinder amplitude with a magnetic field turned on in the
$2,3$-direction, $F_{23}={\cal B} Q$, where $Q$ is an element of the
Cartan subalgebra. This has a trivial effect on the computation; the
effect of the magnetic field\cite{absou} is to shift the oscillator
frequencies for the $2,3$-modes, by a factor
$\varepsilon={1\over\pi}\left( \tan^{-1}\pi q_i {\cal B}+\tan^{-1}\pi
q_j {\cal B}\right)$. As a result, the main effect is to modify one of
the $\theta$-functions, and thus the cylinder amplitude becomes
\beqn
A_{0,0'}(F)&=&{V_4\over 2kk'}\int {dt\over t} (8\pi^2\alpha't)^{-1}\sum_{i,j}
{(q_i+q_j){\cal B}\over 2\pi}
\sum_{n,n'}W_{n,n'}\ a_{n,n'}\ (\gamma_{(n,n')})_{ii}
(\gamma_{(n,n')}^{-1})_{jj}\nonumber\\
&\times &\left[ {\theta_3(i\varepsilon t)\theta_3(z)\theta_3(z-z')\theta_3(z')
-(3\leftrightarrow 2)
-(3\leftrightarrow 4)
+(3\leftrightarrow 1)\over
\theta_1(i\varepsilon t) \theta_1(z)\theta_1(z-z')\theta_1(z')}\right]
\eeqn
Expansion of this formula for small $\varepsilon$ gives us
\beq
A_{0,0'}(F)\sim {v_4\over kk'} \int {d\ell\over\ell} \left( \sin^2\pi z\ 
\tr\gamma_{(z,0)}\ \tr\gamma^{-1}_{(z,0)}F^2 + \ldots\right)+\ldots
\eeq
where the first ellipsis denotes the other twist sectors which contribute.
Note that we may write
 $\tr\gamma^{-1}_{(z,0)}F^2=\sum_{r,s}e^{-2\pi izr} \tr F_{r,s}^2$,
and so the amplitude reduces to
\beq
A_{0,0'}(F)\sim v_4 \int {d\ell\over\ell}\sum_{r,s}\beta_{r,s}\ \tr F_{r,s}^2
+\ldots\eeq
This result clearly demonstrates the gauge coupling renormalization, and is
in agreement with the above discussion.

\subsection{Bending}

We may now relate the above effect directly to bending of the NS-branes
in the brane box theory. First note that one may think of the 3-branes
of the orbifold model as higher dimensional D-branes wrapped on shrunken
cycles at the orbifold points \cite{wrapped}. In the present case, we do
not have a K3 orbifold, but rather a Calabi-Yau orbifold, and so there
are D5-branes wrapped on shrunken 2-cycles, as well as D7-branes wrapped
on shrunken 4-cycles. In this general case, the corresponding effects
discussed in the previous section are not entirely geometric in the
brane box picture.

The situation simplifies however in the ``N=2 limits'', (see also discussion in
Ref. \cite{brand}) corresponding to
confining ourselves to a given twist sector, say $(n,0)$. In this case, we
effectively have an $A$-type singularity (times a $T^2$). In this case,
we have
\beq
\int_{\RRs^{\ 4}} C_{(4)}\equiv \int_{\RRs^{\ 4}\times S^2} C_{(6)}
\eeq
and 
\beq
\int_{\RRs^{\ 4}} C_{(0)}\wedge F\wedge F\equiv \int_{\RRs^{\ 4}\times S^2} C_{(2)}
\wedge F\wedge F
\eeq
Here, $C_{(2)}=B_{NS}+iB_{RR}$. In this simplified case, these Wilson lines
are T-dual to position of the NS5-branes\cite{hanura}. Therefore the $C^{(z,0)}$
and $C^{(0,z')}$
have a geometric interpretation in the brane box theory as bending of 
horizontal and vertical NS5-branes. However, in general the $\beta$-functions
$\beta_{r,s}$ depend on all the Fourier modes $C^{(n,n')}$, most of which
are unrelated to simple bending in the brane box picture.

{\bf Acknowledgments:} 
We thank A. Hanany and Y. Shirman for conversations.
Research supported in part by 
the United States Department of Energy grant DE-FG02-91ER40677. 
RGL is supported by a DOE Outstanding Junior
 Investigator Award.

%%%%%%%%%%%%%%%%%%%%%%%%%%%%%%%%%%%%%%%%%%%%%%%%%%%%%%%%%%%%%%%%%%%%%

\def\npb#1#2#3{Nucl. Phys. {\bf B#1} (#2) #3}
\def\plb#1#2#3{Phys. Lett. {\bf #1B} (#2) #3}
\def\prd#1#2#3{Phys. Rev. {\bf D#1} (#2) #3}
\def\prl#1#2#3{Phys. Rev. Lett. {\bf #1} (#2) #3}
\def\cqg#1#2#3{Class. Quantum Grav. {\bf #1} (#2) #3}
\def\mpl#1#2#3{Mod. Phys. Lett. {\bf A#1} (#2) #3}
\def\hepth#1#2#3#4#5#6#7{hep-th/#1#2#3#4#5#6#7}
\def\jhep#1#2#3{J. High Energy Phys. {\bf #1} (#2) #3}

\end{document}